\documentclass[cits]{PoS}
\usepackage{amstext}
\usepackage{relsize}
\usepackage{booktabs}
\usepackage{multirow}
\usepackage{xspace}

\makeatletter
\DeclareRobustCommand\bfseries{%
  \not@math@alphabet\bfseries\mathbf
  \fontseries\bfdefault\selectfont\boldmath}

\makeatother

\title{Experimental Review on Lepton Universality and Lepton Flavour Violation tests at the B-factories}

\ShortTitle{Lepton Universality and Lepton Flavour Violation tests at the B-factories}

\author{\speaker{Alberto Lusiani}%
         \thanks{Presented for the \babar collaboration.}\\
        INFN and Scuola Normale Superiore -- Pisa\\
        E-mail: \email{alberto.lusiani@pi.infn.it}}

\abstract{Since 1999, the B-factories collaborations \babar and Belle
have accumulated and studied large samples of tau lepton pairs.  The
experimental results on Lepton Universality checks and Lepton Flavour
Violation searches are reported.}

\FullConference{Kaon International Conference\\
                 May 21-25, 2007\\
                 Laboratori Nazionali di Frascati dell'INFN}

\graphicspath{{figures/}{../figures/}}

\newcommand{\babar}{\mbox{%
    \slshape B\kern-0.1em{\smaller A}\kern-0.1em
    B\kern-0.1em{\smaller A\kern-0.2em R}}\xspace}

\newcommand{\invfb}{\ensuremath{\,\text{fb}^{-1}}\xspace}
\newcommand{\fs}{\ensuremath{\,\text{fs}}\xspace}
\newcommand{\EE}[1]{\ensuremath{\cdot 10^{#1}}}
\newcommand{\etal}{\textit{et al.}\xspace}
\newcommand{\prelim}{\ensuremath{^*}}

\def\BR{{\ensuremath{\cal B}}\xspace}

\begin{document}

\section{Introduction}

In recent years, the \babar and Belle experiments have contributed
results on tau lepton physics, which improved the experimental picture
of lepton universality and of lepton flavor violation searches.

Both experiments rely on ``B-factories'' operating at a centre-of-mass
energy of 10.58\,GeV on the $\Upsilon(4s)$ peak, just above the
threshold for producing $B$-mesons. 
\babar operates at the PEP-II complex at SLAC, which collides 9\,GeV
electrons against 3.1\,GeV positrons, and has recorded about 420\invfb
of data by May 2007. Belle operates at the KEKB B-factory in Japan,
which collides 8\,GeV electrons against 3.5\,GeV positrons, and has
recorded about 710\invfb of data by May 2007.

The \babar~\cite{ref:babarDet} and Belle~\cite{ref:belleDet} detectors
share several similarities and both include a silicon microvertex
detector, a drift chamber, a 1.5\,T solenoidal superconducting magnet,
an electromagnetic calorimeter based on Cesium Iodide crystals, and a
segmented muon detector in the magnet return yoke.  The two
experiments differ in the particle identification strategy: Belle uses
an aerogel threshold Cherenkov detector together with time-of-flight
and tracker dE/dx, whereas \babar relies on a ring-imaging Cerenkov
detector supplemented by the dE/dx in the trackers.

With a total now exceeding 1.1\,ab$^{-1}$ of integrated luminosity and
a $e^+e^-\to\tau^+\tau^-$ cross-section at 10.58\,GeV of
0.919\,nb~\cite{Banerjee:2007is}, B-factories recorded in excess of
$10^9$ tau pairs, which allow for improving statistics-limited
results, like in particular searches for tau lepton flavor violating
decays.

\section{Lepton universality tests}

The Standard Model (SM) predicts that all lepton doublets have
identical couplings to the $W$ boson.  Ratios of measured decay widths
of leptonic or semileptonic decays which only differ in the lepton
flavour test whether the $W$ interaction is universal for the three
lepton flavours. The present data are consistent with the universality of
the leptonic charged-current couplings to the 0.2\%
level~\cite{aPichTau06}. B-factories have produced results on the tau
mass, the tau lifetime, and could in principle measure tau branching
fractions that improve the experimental knowledge on the less
precisely known factors in the following expressions for leptonic
coupling ratios, which the SM predicts to be consistent with unity:
\begin{eqnarray}
  \frac{\Gamma_{\tau\to e}}{\Gamma_{\mu\to e}}
  \;\propto\;
  \left(\frac{g_\tau}{g_\mu}\right)^2
  &=&
  \frac{\tau_\mu}{\tau_\tau}%
  \BR(\tau^- \to e^-{\bar \nu_e}\nu_\tau)
  \left(\frac{m_\mu}{m_\tau}\right)^5
  \frac{f(m^2_e/m^2_\mu)r^\mu_{EW}}{f(m^2_e/m^2_\tau)r^\tau_{EW}}
  \label{ratio1}
  \\
  \frac{\Gamma_{\tau\to\mu}}{\Gamma_{\mu\to e}}
  \;\propto\;
  \left(\frac{g_\tau}{g_e}\right)^2
  &=&
  \frac{\tau_\mu}{\tau_\tau}%
  \BR(\tau^- \to \mu^-{\bar \nu_\mu}\nu_\tau)
  \left(\frac{m_\mu}{m_\tau}\right)^5
  \frac{f(m^2_e/m^2_\mu)r^\mu_{EW}}{f(m^2_\mu/m^2_\tau)r^\tau_{EW}}
  \label{ratio2}
  \\
  \frac{\Gamma_{\tau\to e}}{\Gamma_{\tau\to\mu}}
  \;\propto\;
  \left(\frac{g_e}{g_\mu}\right)^2
  &=&
  \frac{%
    \BR(\tau^- \to e^-{\bar \nu_\mu}\nu_\tau)}{%
    \BR(\tau^- \to \mu^-{\bar \nu_\mu}\nu_\tau)}
  \frac{f(m^2_\mu/m^2_\tau)}{f(m^2_e/m^2_\tau)}
  \label{ratio3}
\end{eqnarray}
\noindent In the above expressions, $\Gamma$ represents partial
widths, $\tau_\ell$ lepton lifetimes, $m_\ell$ lepton masses,
\BR branching fractions; $f(x) = 1 - 8x + 8x^3 - x^4 -12x \ln
x$ are phase space factors~\cite{ref:marc88},
$r^\ell_{EW} \approx 1$ correspond to electro-weak radiative
corrections~\cite{ref:marc88}.

\subsection{Tau mass}

The mass of the tau lepton appears at the fifth power in the coupling
ratio expressions~\ref{ratio1} and \ref{ratio2}, contributing a relative
uncertainty of $0.08\%$~\cite{PDG}.

BELLE measured the tau mass using a pseudomass technique that was
first employed by the ARGUS collaboration~\cite{ARGUS}.  Tau pairs
candidates are selected where one tau decays into a single prong
(electron or muon), and the other one decays into 3-prongs, all of
which are pions, with no additional $\pi^0$. Then a quantity
$M_{\text{min}}$ is computed:
\begin{equation}
 M_{\mathrm{min}}=\sqrt{M_X^2+2(E_{\mathrm{beam}}-E_X)(E_X-P_X)},
\end{equation}
which is less than or equal to the tau lepton mass.  $M_X$, $E_X$ and
$P_X$ are the invariant mass, energy and absolute value of the
momentum, respectively, of the hadronic system in the center-of-mass
(c.m.) frame, and
$E_{\mathrm{beam}}$ is the energy of the electron (or positron) in
this frame. The distribution of $M_{\mathrm{min}}$ extends up to and
has a sharp edge at $M_\tau$, smeared by detector resolution and
initial and final state radiation. The edge position obtained from a
fit to the $M_{\mathrm{min}}$ distribution is relatively insensitive
to the background and is a precise estimator of the tau mass.  The
offset between the edge and the tau mass has to be estimated with a
Monte Carlo. The measurement (on 414\invfb of data) reaches a
precision close to the present BES-dominated world average, $m_\tau =
1776.99\,{}^{+\, 0.29}_{-\,0.26}$~\cite{PDG}:
\begin{equation}
  m_\tau = 1776.61\pm 0.13\pm 0.35\,\text{MeV}\quad
  \text{\cite{ref:belleTauMass}}.
\end{equation}

\begin{figure}[tb]
\begin{center}
\includegraphics[width=0.5\linewidth,clip]{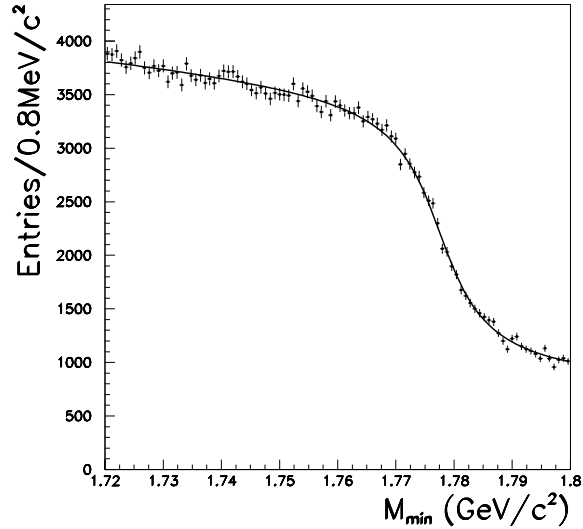}
\caption{Pseudomass distribution for $M_{\text{min}}$ for the
  $\tau\to3\pi\nu$ candidates. The tau mass is a parameter of the fit
  shown with a solid line.}
\label{fig:belleTauMass}
\end{center}
\end{figure}

The limiting systematic contributions to this result come from the
understanding of the momentum scale and the reliability of the
simulation in estimating the offset between the tau mass and the edge
position.  The uncertainty on the momentum scale is
assessed by comparing the reconstructed $B$ meson masses to their
world average, which is know up to about
$1.5\,\mathrm{MeV}$~\cite{PDG},
and noting that the di-muon invariant mass peak matches
the center of mass event energy determined by the beam energies
withing $3\,\mathrm{MeV}$.  Apparently, measurement conducted at
threshold with machines where the beam energy can be calibrated
through resonant depolarization, suffer from smaller systematic
uncertainties, as the recent KEDR result shows
($1776.80^{{+}0.25}_{{-}0.23} \pm 0.15~\mathrm{MeV}$~\cite{KEDR}).
KEDR aims to obtain a final accuracy of 0.15\,MeV, and BESIII aims at
a precision better than 0.1\,MeV.

\subsection{Tau lifetime}

The tau lepton lifetime is known up to 0.3\% and is the least
precisely known factor in the coupling ratio expressions~\ref{ratio1}
and \ref{ratio2}.  \babar has presented a preliminary measurement of
the tau lifetime with an error comparable to the present world
average:
\begin{equation}
  \tau_\tau = 289.4 \pm 0.9 \pm 0.9\,\fs\quad
  \text{\cite{ref:babarTauLife}}.
\end{equation}
The measurement uses about 80\invfb of data and is based on an
extremely pure (99.4\%) yet scarcely efficient (0.2\%) selection of 1
against 3-prong events in the c.m.\ system, where the 1-prong track is
an identified electron.  Electron identification is used because it is
more efficient and less contaminated with hadrons with respect to muon
tagging. The selected tau candidates are about 300,000 and include
about 0.2\% hadronic background, 0.4\% Bhabha background and a
negligible amount of two-photon events.

The measurement is based on the reconstruction of the decay length of
the tau that decayed into the 3-prong tracks. Using a novel technique
aimed at minimizing the systematic dependence on the detector
alignment, the tau decay vertex is first computed in a plane
transverse with respect to the beam axis. The transverse decay length
is computed within the transverse plane by projecting the vector from
the luminous region center to the tau decay vertex along 3-prong total
momentum direction (which approximates the tau flight direction).  The
tau decay length is finally reconstructed by projecting the transverse
decay length onto the 3-prong total momentum direction.

The mean decay length is determined with an average, abstaining on
purpose from weighting events according to their estimated errors in
order to minimize systematic effects from alignment and detector
material modeling.  The mean lifetime is determined using the Monte
Carlo prediction of the average tau momentum, using the KKMC
generator~\cite{KKMC}, which includes complete 2nd order radiative
corrections. The measurement offset that originates from tracking
errors correlations~\cite{sysbiases} and from approximating the tau
momentum direction with that of the 3-prong total momentum is
subtracted using the Monte Carlo simulation.

Finally, the contribution of background is subtracted.  While hadronic
background is simulated, a statistically adequate Monte Carlo
simulation of Bhabha events is impractical because the relevant
cross-section is about 20 times the tau production cross section
($\approx 1\,\text{nb}$), therefore a data control sample is used to
estimate both the Bhabha contamination and its decay length
distribution.

Systematic uncertainties come mainly from the reliability of the
measurement bias subtraction using Monte Carlo, from detector
alignment, from the mean tau momentum Monte Carlo simulation, and from
background subtraction. This measurement includes a study of
the effects of detector misalignment. Decay length shifts with respect
to a perfectly aligned detector are measured on simulated Monte Carlo
events by refitting tracks from coordinates taken on a detector that
is purposefully distorted. Six representative distortions are applied
by displacing the silicon vertex detector wafer positions according to
the observed distortions and uncertainties that are derived from
reconstructed data.

\begin{figure}[tb]
\begin{center}
\begin{tabular}{@{}cc@{}}
\fbox{\includegraphics[width=0.47\linewidth,clip]{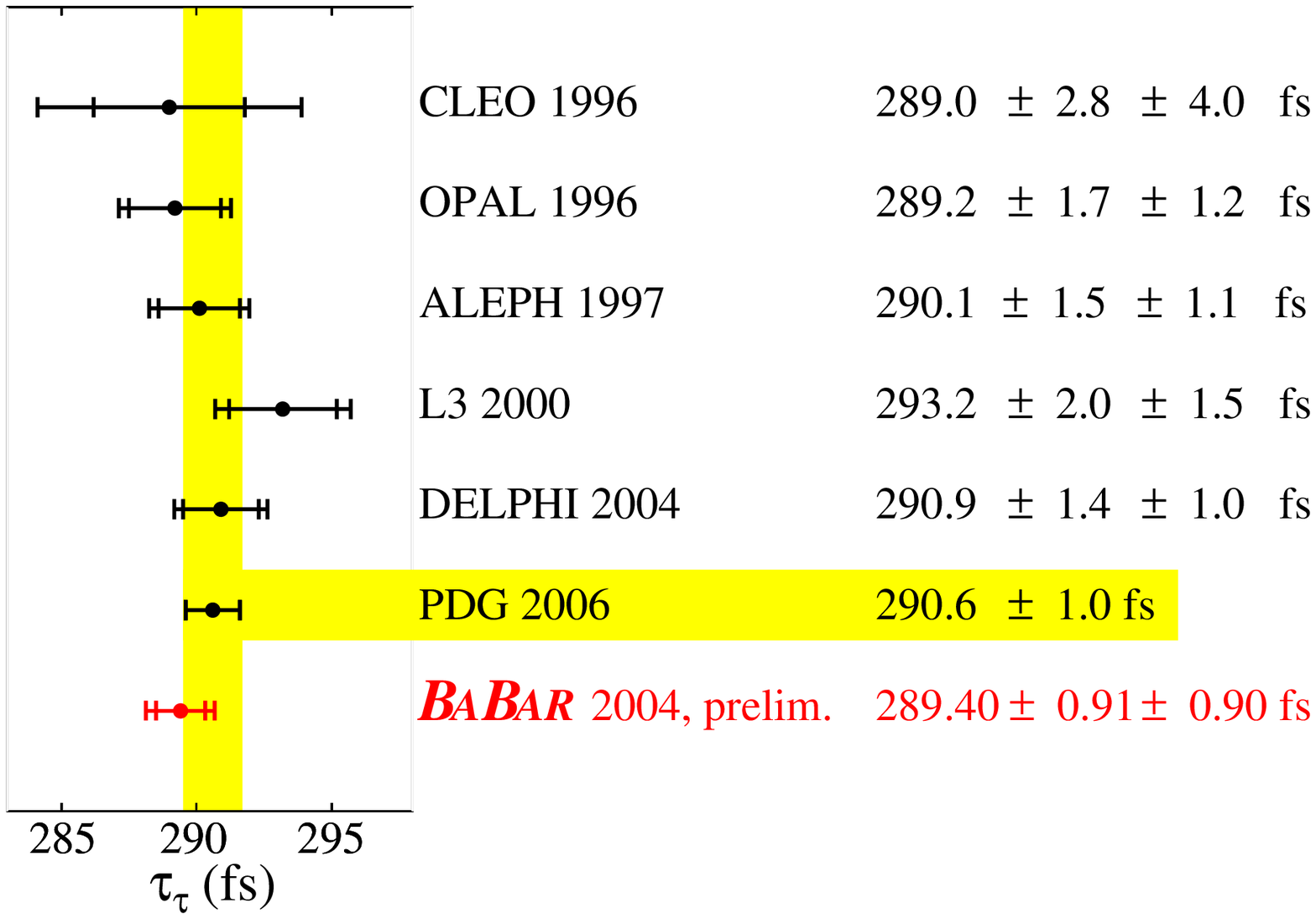}} &
\fbox{\includegraphics[width=0.47\linewidth,clip]{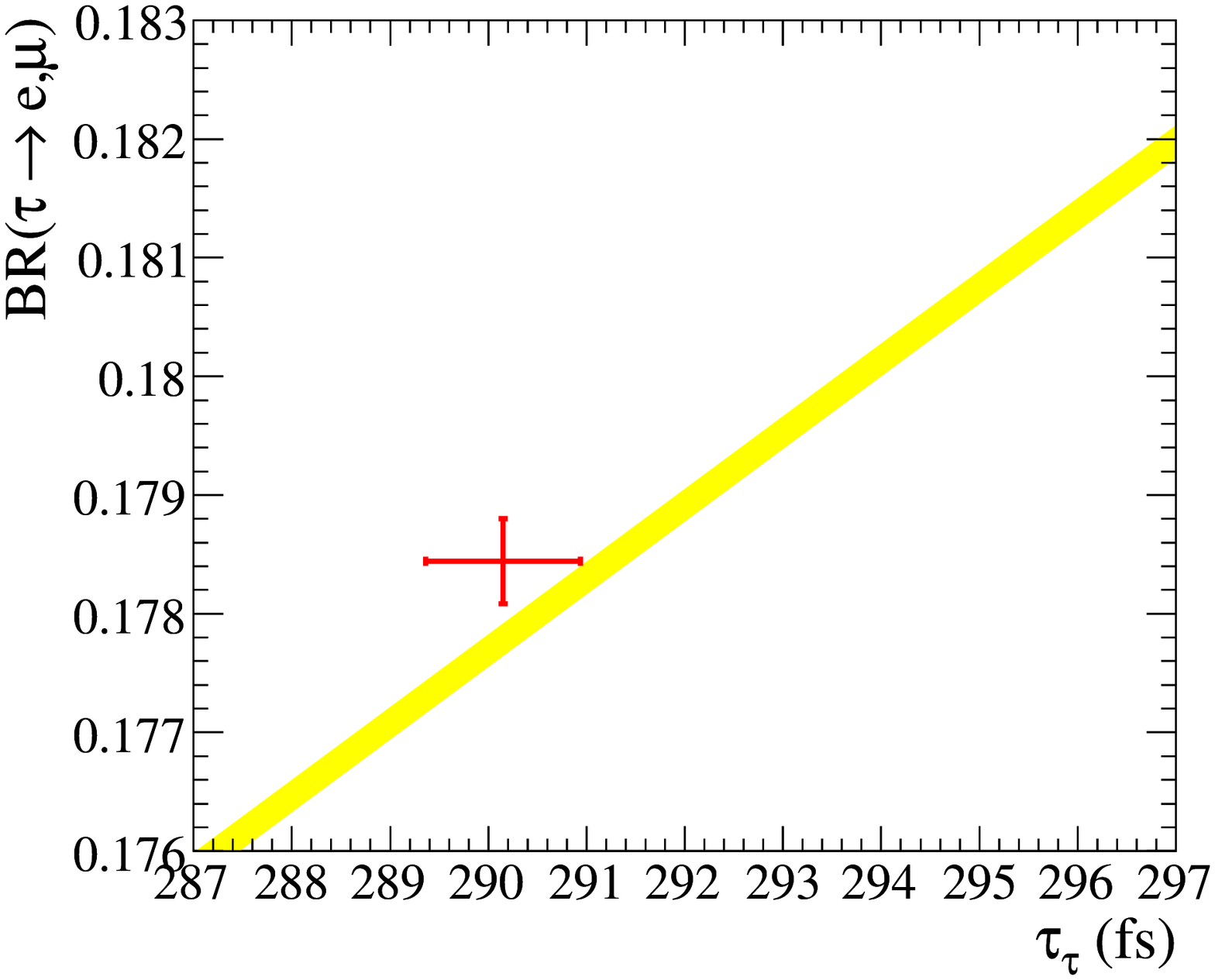}}
\end{tabular}
\caption{Selected tau lifetime measurements (left) and check of the
  Standard model (SM) prediction of universal leptonic couplings to the
  $W$ (right) combining the present tau lifetime world average with
  the \babar preliminary result. The thickness of the oblique line
  represents the uncertainty of the SM constraint, and is dominated by
  the uncertainty on the tau mass.}
\label{fig:babarTauLife}
\end{center}
\end{figure}

This preliminary measurement contributes the largest experimental
improvement in recent years for the coupling ratio
expressions~\ref{ratio1} and \ref{ratio2}.
Figure~\ref{fig:babarTauLife} reports the measurement compared with
the present world average and previous selected
measurements~\cite{ref:tauLifeAll}.  Combining the \babar 2004 result with the
present world average assuming no systematic error correlations we
obtain
\begin{equation}
 \tau_\tau = 290.15 \pm 0.79\,\fs.
\end{equation}
Using the present world averages, we present an updated check of
lepton universality in Figure~\ref{fig:babarTauLife} and updated
determinations of the coupling ratios
\begin{eqnarray}
  \frac{g_\mu}{g_\tau} = 0.9982 \pm 0.0020, \quad
  \frac{g_e}{g_\tau} = 0.9980 \pm 0.0020, \quad
  \frac{g_{e,\mu}}{g_\tau} = 0.9981 \pm 0.0017,
\end{eqnarray}
where $g_{e,\mu}$ is determined assuming $g_e = g_\mu$ holds for the
theory.

In the recent past, LEP experiments improved considerably the
experimental precision on the tau lifetime profiting from ideal
conditions in most respects but statistics: there high momentum tracks
had small impact parameter errors due to multiple scattering, tau
events had a distinctive topology that permitted a pure and efficient
selection against backgrounds, vertex detectors provided precise
tracking close to the origin and systematic uncertainties from
detector misalignment were reduced thanks to the complete and uniform
acceptance in the azimuthal angle~\cite{sysbiases}.  B-factories
appear to be the only facilities where the tau lifetime measurement
can be improved in the near future and they can overcome with
statistics the disadvantages related to increased multiple scattering
due to lower momenta and to less favourable physics conditions for an
efficient and pure selection.

\subsection{Tau leptonic branching fractions}

B-factories have not yet succeeded in matching LEP experiments
regarding the measurement of the tau leptonic branching fractions,
because the systematic precision is limited by uncertainties on the
normalization of the number of produced tau leptons, arising from
uncertainties on the integrated luminosity corresponding to the
analyzed event samples and on the cross-section for tau pair
production at the $\Upsilon(4s)$. Uncertainties on luminosity are
about 1\% to be compared with better than 0.1\% for LEP experiments.
Until recently, the uncertainty on the tau pair cross-section was
estimated by comparing the KoralB~\cite{ref:KoralB} and KKMC
predictions to be $2.2\%$. However there has been remarkable progress
that led to a new estimate of the tau pair cross section at the
$\Upsilon(4s)$:
\begin{equation}
  \sigma(e^+e^-\to\tau^+\tau^-) = 0.919 \pm 0.003\,\text{nb}\quad
  \text{\cite{Banerjee:2007is}}.
\end{equation}
Furthermore, B-factories may measure ratios of branching fractions
(see eq.~\ref{ratio3}) whose uncertainties won't be limited by the
understanding of luminosity and cross-section, but only by the
systematics related to electron versus muon identification and to
background suppression and subtraction.

\section{Lepton Flavour Violation tests at the B-factories}

\subsection{Common analysis features}

The typical tau LFV decay search at the B-factories selects low track
multiplicity events that have 1 against 1 or 3 tracks in the c.m.\
frame. The thrust axis is used to define two hemispheres, each
of which is then examined for consistency with a tau LVF decay, while
the other one must be compatible with a known tau decay.
Unlike known tau decays, which include at least one neutrino, the
reconstructed products of a LFV tau decay are expected to match the
tau mass and half the c.m. energy within the experimental
resolution.  It is worth noting that physics effects also limit the
experimental accuracy in reconstructing the parent tau energy and mass
from its decay products: initial and final state radiation affect the
tau energy itself before decay, and radiation in decay and
Bremsstrahlung from the decay products change the reconstructed
energy and invariant mass.
The energy is reconstructed with a typical resolution of 50\,MeV and,
when using a total energy constraint to half the c.m.\ energy, the
invariant mass is reconstructed with a resolution of about 10\,MeV.
Selected events around the expected energy and mass within 2 or 3
standard deviations are then investigated looking for an
excess over the expected background.

The amount of expected background is normally estimated using the
distribution shapes from the Monte Carlo simulation normalized to the
observed events in a two-dimensional sideband region around the signal
region on the energy-mass plane. The signal efficiency is estimated
with a Monte Carlo simulation and typically lies between 2\% and 10\%
depending on the channel. Typical cumulative efficiency components
include 90\% for trigger, 70\% for geometrical acceptance and
reconstruction in the detector, 70\% for reconstructing the selected
track topology, 50\% for particle identification, 50\% for additional
selection requirements before checking the reconstructed energy and
mass, and 50\% for requiring consistency with the expected energy and
mass. The selection efficiency and background suppression are
optimized to give the best "expected upper limit" assuming that the
data contain no LFV signal. The optimization and all systematic
studies are completed while maintaining the experimenter ``blind'' to
data events in the signal box in the energy-mass plane, in order to
avoid experimenter biases. 

When the expected background in the signal region is of order one or
less, the number of signal events is normally set to the number of observed
events minus the background, while in presence of sizable background
the numbers of background and signal events are concurrently determined
from a fit to the mass distribution of events that have total energy
compatible with the expected one.

\subsection{Results}

LFV decays can be grouped in the following categories: tau to 
lepton-photon ($\tau\to\ell\gamma$, where $\ell=e,\mu$), tau to three
leptons or one lepton and two charged hadrons
($\tau\to\ell_1\ell_2\ell_3$, $\tau\to\ell h_1 h_2$), tau to a lepton
and a neutral hadron ($\tau\to\ell h^0$, where $h^0 = \pi^0, \eta,
\eta^\prime, K_s^0, \text{\textit{etc.}}$).

Belle has reported preliminary results for $\tau\to\mu\gamma$ and
$\tau\to e\gamma$~\cite{Hayasaka:2007vc} based on the analysis of
$535\invfb$ of data. The $\tau\to\mu\gamma$ ($\tau\to e\gamma$)
searches have a 5.1\% (3\%) signal efficiency within a $2\sigma$
elliptical signal
region in the energy-mass plane plane. A two-dimensional unbinned
extended maximum likelihood fit for signal and background in the
signal region obtains $-3.9^{+3.6}_{-3.2}$ ($-0.14^{+2.18}_{-2.45}$)
signal and $13.9^{+6.0}_{-4.8}$ ($5.14^{+3.86}_{-2.81}$) background
events.  Frequentist 90\%\,CL limits are obtained by running Monte
Carlo simulations in which the signal is increased until 90\% of the
fits obtain more than the observed signal events:
$\BR(\tau\to\mu\gamma) < 0.45\EE{-7}$ and
$\BR(\tau\to e\gamma) < 1.2\EE{-7}$.

\babar has recently submitted for publication~\cite{Aubert:2007pw} and
Belle has recently reported~\cite{Abe:2007ev} improved results on tau
to three leptons LFV searches based on enlarged event samples of
$400\invfb$ and $535\invfb$ respectively.  The two analyses have
similar signal efficiencies ($5.5\%{-}12.5\%$) although different
strategies are adopted to define the signal regions in the energy-mass
plane: Belle uses ellipses large enough to contain 90\% of signal
events, while \babar uses rectangular signal boxes
optimized to obtain the lowest expected upper limit in case there is
no signal.  The Belle selection is significantly more effective in
suppressing the background, which is expected to be $0.01{-}0.07$
events in all channels but the three electron one, where 0.4 events
are expected because of Bhabha contamination. The
background in the signal region is estimated from the mass
distribution sidebands, assuming it is constant, using looser
selection criteria to get reasonable samples close to the signal
region. No details are given on how the extrapolation is done for the
full selection.  \babar expects $0.3{-}1.3$ background events, i.e.\
of order one, since the selection is optimized for the
best expected upper limit, at the risk of getting background events in
the signal region.  Belle observes no candidate signal events in
$535\invfb$ of data in all modes, and calculates upper limits
using Feldman and Cousin
ordering~\cite{Abe:2007ev,ref:pole} in the range
$[2.0{-}4.1]\EE{-8}$, depending on the mode.  \babar observes from 0 to
2 events in $376\invfb$ of data, and calculates upper limits according
to Cousin and Highland prescription~\cite{ref:CousinHighland} with no
Feldman and Cousin ordering in the range $[4-8]\EE{-8}$.

\babar has recently published new results on tau LFV decays into a
lepton and a hadron pseudoscalar
$\pi^0,\eta,\eta^\prime$~\cite{Aubert:2006cz}.
In these analyses both of the $\eta\to\gamma\gamma$ and the $\eta\to
3\pi$ decay modes are used for the analyses, and $\eta^\prime$
candidates decaying both to $\eta 2\pi$ and $\gamma 2\pi$ are
considered.
The expected background per channel is between 0.1 and 0.3 events.
Summing over all ten modes, 3.1 background events are expected, and 2
events are observed.

Belle has recently reported improved results on tau LFV decays into a
lepton and a vector meson $V^0$ ~\cite{Abe:2007ex}, with $V^0 =$
$\phi$, $\omega$, $K^{*0}$ or $\bar{K}^{*0}$, using 543\invfb of
data. No excess of signal events over the expected background is
observed, and upper limits the branching fractions are obtained in the
range $(0.7{-}1.8)\EE{-7}$ at 90\%\,CL.

\babar reported also on less conventional searches of tau LFV decays
into $\Lambda\pi$~\cite{Aubert:2006uu} and of LFV in tau production
($e^+e^-\to\ell\tau$)~\cite{Aubert:2006uy}, finding no signal.

The above results and additional B-factories LFV
results~\cite{Abe:2006qv,Aubert:2005tp,Miyazaki:2006sx} are
summarized in Table~\ref{tab:lfvSearches}.
At the Tau2006 conference in Pisa, Swagato Banerjee presented
frequentist combinations~\cite{Banerjee:2007rj} of some measurements,
which are also included in the table.

\begin{table}[tb]
  \begin{center}
    \begin{tabular}{lcccccc}\toprule
        \multirow{2}[0]{*}{Channel}
        & \multicolumn{2}{c}{Belle}
        & \multicolumn{2}{c}{\babar}
        & \multicolumn{2}{c}{combined}
        \\ \cline{2-7}

        & UL90 & Lumi 
        & UL90 & Lumi 
        & UL90 & Lumi 
        \\

        & $(10^{-7})$ & (fb$^{-1}$)
        & $(10^{-7})$ & (fb$^{-1}$)
        & $(10^{-7})$ & (fb$^{-1}$)
        \\ \hline

        $\mu\gamma$ & 0.5\prelim & 535 & 0.7 & 232 & 0.16 & 767 \\
        $e\gamma$ & 1.2\prelim & 535 & 1.1 & 232 & 0.94 & 767 \\
        $\mu\eta$ & 0.65\prelim & 401 & 1.5 & 339 & 0.51 & 740 \\
        $\mu\eta^\prime$ & 1.3\prelim & 401 & 1.3 & 339 & 0.53 & 740 \\
        $e\eta$ & 0.92\prelim & 401 & 1.6 & 339 & 0.45 & 740 \\
        $e\eta^\prime$ & 1.6\prelim & 401 & 2.4 & 339 & 0.90 & 740 \\
        $\mu\pi^0$ & 1.2\prelim & 401 & 1.5 & 339 & 0.58 & 740 \\
        $e\pi^0$ & 0.8\prelim & 401 & 1.3 & 339 & 0.44 & 740 \\
        $\ell\ell\ell$ & $0.20{-}0.41$\prelim & 535 & $0.4{-}0.8$\prelim & 376 \\
        $\ell h h^\prime$ & $2{-}16$ & 158 & $1{-}5$ & 221 \\ 
        $\ell V^0$ & $0.7{-}1.8$\prelim & 543 &  &  \\

        $\mu K_S$ & 0.49 & 281 & & \\
        $e K_S$ & 0.56 & 281 & & \\
        $\Lambda\pi,\overline\Lambda\pi$ & & & $5.8{-}5.9$\prelim & 237 \\
        $\Lambda K, \overline\Lambda K$ & & & $7.2{-}15$\prelim & 237 \\
        $\sigma_{\ell\tau}/\sigma_{\mu\mu}$ & & & $40{-}89$ & 211 \\
        \bottomrule
        \multicolumn{5}{l}{($^*$ preliminary)}
      \end{tabular}
  \end{center}
  \caption{Summary of 90\%\,CL upper limits on tau LFV decays from the
    B-factories. An asterisk indicates a preliminary result. $h$ and
    $h^\prime$ denote a charged pion or kaon. Banerjee's combination of
    a subset of these channels is also included.}
  \label{tab:lfvSearches}
\end{table}

\subsection{Prospects}

While Belle plans to run until it will collect 1\,ab$^{-1}$ of data,
\babar is funded to run until September 2008, when it expects to reach
a total integrated luminosity of about 0.8\,ab$^{-1}$. In case there
is no signal, the expected upper limits on the number of selected
signal events will improve depending of the amount of irreducible
background in each channel:
\begin{itemize}

\item
  when the expected background is large ($N_{\text{BKG}} \gg 1$), the
  expected upper limit is
  $N^{\text{UL}}_{90} \approx 1.64\sqrt{N_{\text{BKG}}}$;

\item
  when the expected background is small ($N_{\text{BKG}} \ll 1$),
  using~\cite{ref:CousinHighland} one gets $N^{\text{UL}}_{90} \approx
  2.4$.

\end{itemize}
Reducing the background below few events does not much improve the
expected limit if significant efficiency is lost in the process,
therefore optimized searches often enlarge the acceptance until
$N_{\text{BKG}} \approx 1$.  For the cleaner channels,
analyses can be optimized for an increased data sample to
keep $N_{\text{BKG}} \approx 1$ without loosing a significant part of
the signal efficiency: in this best case scenario, the expected upper
limits will scale as $N_{\text{BKG}}/{\cal L}$ i.e.\ as $1/{\cal L}$.
On the other hand, if no optimization is possible, just keeping the
current analyses will provide upper limits that scale as
$\sqrt{N_{\text{BKG}}}/{\cal L}$, i.e. as $1/\sqrt{{\cal L}}$.

Mike Roney has recently estimated~\cite{RoneyMoriond07} that for
$\tau\to\ell\gamma$ analyses the background coming from
$\tau\to\ell\nu\nu$ decays and initial and final state photons can be
considered ``irreducible'' at its present level (20\% of the
total).  In these conditions, the final combined Belle and \babar data
set will allow for expected upper limits in the range
$[0.1{-}0.2]\EE{-7}$.  The actual combined upper limit obtained by
Banerjee~\cite{Banerjee:2007rj} for $\tau\to\mu\gamma$ is already in
that range as a consequence of a downward fluctuation of the observed
events with respect to the expected background.

Other tau LFV decay channels ($\tau\to\ell\ell\ell$, $\tau\to\ell hh$,
$\tau\to\ell h^0$) do not yet appear to be background limited and
their expected expected upper limits with the full B-factories dataset
are about $0.1\EE{-7}$.

Beyond the current B-factories facilities, there are proposals for
super B-factories~\cite{superb} that would permit a 100 fold increase
in the size of the tau pairs sample: this would allow probing tau LFV
decays at the $10^{-9}{-}10^{-10}$ level.

\acknowledgments

I am grateful to Hisaki Hayashii for his assistance in getting the
Belle experimental results.

\newif\ifbibtitles
\bibtitlesfalse
\bibtitlestrue
\newcommand{\bibvnbf}[1]{{\bf #1}}
\newcommand{\bibpy}[2]{ (#2) #1}
\newcommand{\jprBase}        {Phys.\ Rev.\xspace}
\newcommand{\jprd}      [1]  {\jprBase\ D~\bibvnbf{#1}}

\end{document}